\documentclass[twoside,draft]{article}
\usepackage{latexsym,e-journal}

\usepackage[letterpaper]{geometry}
\usepackage[utf8]{inputenc}
\usepackage[T1]{fontenc}
\usepackage[english]{babel}

\usepackage{amssymb,amsfonts,amsmath}

\usepackage[perpage,symbol*]{footmisc}
\usepackage[final]{graphicx}
\usepackage{pstricks}
\usepackage{cite}

\usepackage[varg]{txfonts}

\begin{document}

\renewcommand{\refname}{References}
\renewcommand{\tablename}{\small Table}
\renewcommand{\figurename}{\small Fig.}
\renewcommand{\contentsname}{Contents}

\def \pteptitle {Something is wrong in the state of QED}
\def \ptepauthor {Oliver Consa}

\twocolumn[%
\begin{center}
\renewcommand{\baselinestretch}{0.93}
{\Large\bfseries \pteptitle}\par
\renewcommand{\baselinestretch}{1.0}
\bigskip
Oliver Consa\\ 
{\footnotesize  Department of Physics and Nuclear Engineering, Universitat Politècnica de Catalunya \\ 
Campus Nord, C. Jordi Girona, 1-3, 08034 Barcelona, Spain\rule{0pt}{8pt}\\
E-mail: oliver.consa@gmail.com
}\par
\medskip
{\small\parbox{11cm}{%
Quantum electrodynamics (QED) is considered the most accurate theory in the history of science. However, this precision is based on a single experimental value: the anomalous magnetic moment of the electron (g-factor). An examination of the history of QED reveals that this value was obtained in a very suspicious way. These suspicions include the case of Karplus \& Kroll, who admitted to having lied in their presentation of the most relevant calculation in the history of QED. As we will demonstrate in this paper, the Karplus \& Kroll affair was not an isolated case, but one in a long series of errors, suspicious coincidences, mathematical inconsistencies and renormalized infinities swept under the rug. 
}}\smallskip
\end{center}]{%

\setcounter{section}{0}
\setcounter{equation}{0}
\setcounter{figure}{0}
\setcounter{table}{0}
\setcounter{page}{1}

\markboth{\ptepauthor. \pteptitle}{\ptepauthor. \pteptitle}

\markright{\ptepauthor. \pteptitle}
\section{Introduction}
\markright{\ptepauthor. \pteptitle}

After the end of World War II, American physicists organized a series of three transcendent conferences for the \linebreak development of modern physics: Shelter Island (1947), \linebreak Pocono (1948) and Oldstone (1949). These conferences were intended to be a continuation of the mythical Solvay conferences. But, after World War II, the world had changed.

The launch of the atomic bombs in Hiroshima and \linebreak Nagasaki (1945), followed by the immediate surrender of \linebreak Japan, made the Manhattan Project scientists true war heroes. Physicists were no longer a group of harmless intellectuals; they became the powerful holders of the atomic bomb's secrets. They were militarized and their knowledge became a state secret. There was a positive aspect: the US government created the Atomic Energy Commission (AEC) and appointed Oppenheimer as its chief advisor. Former members of the Manhattan Project took control of universities and research centers. They received generous grants from the government that allowed them to invest in expensive experimental resources, such as particle accelerators, supercomputers, or atomic explosion tests. But no one considered the risk to the future of science. 

Former members of the Manhattan Project now enjoyed unlimited credibility. Their hypotheses were automatically accepted and no one could refute his theories. Their calculations and experimental data were subject to military secrecy, and the cost of the equipment necessary to conduct the experiments was prohibitive for the rest of the international scientific community. Consequently, the calculations and experiments could no longer be reproduced independently. Those who accepted the their hypotheses were rewarded with good jobs at research centers and universities, while those who criticized their work were rejected and ostracized. The inevitable consequences of this new situation will soon become apparent. 

\markright{\ptepauthor. \pteptitle}
\section{Quantum Field Theory (QFT)}
\markright{\ptepauthor. \pteptitle}

\subsection{Nature is absurd}
The acceptance of quantum mechanics meant the acceptance of strange explanations, such as the wave-corpuscle duality, the uncertainty principle or the collapse of the wave function.

With the quantization of the electromagnetic field, these strange explanations became even more confusing, including the polarization of the quantum vacuum, electrons and photons interacting with their own electromagnetic fields, particles traveling back in time, the emission and reception of virtual photons, or the continuous creation and destruction of electron-positron pairs in a quantum vacuum.

Feynman summarized this new paradigm quite clearly:\textbf{\textit{ “It is whether or not the theory gives predictions that agree with experiment. It is not a question of whether a theory is philosophically delightful, or easy to understand, or perfectly reasonable from the point of view of common sense. The theory of quantum electrodynamics describes Nature as absurd from the point of view of common sense. And it agrees fully with experiment. So I hope you can accept Nature as She is: absurd. I’m going to have fun telling you about this absurdity, because I find it delightful. Please don’t turn yourself off because you can’t believe Nature is so strange.”}} \cite{Feynmanbook}

\subsection{The problem of infinities}
After the success of the Dirac equation in 1928, quantum mechanics theorists attempted to quantify the electromagnetic field by creating the quantum field theory (QFT). Unfortunately, QFT was a complete failure since any attempted calculation under this theory resulted in an infinite number.

The only solution the proponents could devise was to simply ignore these infinities. Many methods can be used to ignore infinities, but the primary ones are:

\begin{itemize}
\item Substitution: replacing a divergent series with a specific finite value that has been arbitrarily chosen (for example, the energy of an electron).
\item Separation: separating an infinite series into two components, one that diverges to infinity and another that converges to a finite value. Eventually, the infinite component is ignored and only the finite part remains.
\item Cut-off: focusing on an arbitrary term in the evolution of a series that diverges to infinity and ignoring the rest of the terms of the series.
\end{itemize}

All these techniques are illegitimate from a mathematical perspective, as demonstrated by Dirac: \textbf{\textit{“I must say that I am very dissatisfied with the situation because this so-called 'good theory' does involve neglecting infinities which appear in its equations, ignoring them in an arbitrary way. This is just not sensible mathematics. Sensible mathematics involves disregarding a quantity when it is small – not neglecting it just because it is infinitely great and you do not want it!.}} ”\cite{Dirac2}

This technique of ignoring infinities is called renormalization. Feynman also recognized that this technique is not mathematically legitimate: \textbf{\textit{“The shell game that we play is technically called 'renormalization'. But no matter how clever the word, it is still what I would call a dippy process! Having to resort to such hocus-pocus has prevented us from proving that the theory of quantum electrodynamics is mathematically self-consistent. It's surprising that the  theory still hasn't been proved  self-consistent one way or the other by now; I suspect that renormalization is not mathematically legitimate.”}} \cite{Feynmanbook}

As an example of the use of these renormalization techniques we can look at the calculation of the Casimir effect. \cite{Casimir} The equation of the Casimir effect depends on the Riemann function. 
\begin{equation} 
\frac{F_c}{A} = \frac{d}{da}\frac{<E>}{A} =- \frac{\hbar c \pi ^2}{2 a^4} \ \zeta(-3) = \frac{\hbar c \pi ^2}{20 a^4} \ \zeta(-1)
\end{equation} 

However, the Riemann function is defined only for positive values, since for negative values the Riemann function diverges to infinity. The Riemann function of -1 is equal to the sum of all positive integers. Applying a renormalization technique, the Indian mathematician Ramanujan calculated that the sum of all positive integers is not infinity but -1/12. \cite{Integers} 
\begin{equation} 
\zeta(-1) = \sum _{n=1}^{\infty }{n} = 1+2+3+4+5+ ... =  \frac{-1}{12}
\end{equation}
And this is precisely the value that is used in the equation of the Casimir effect.
\begin{equation} 
\frac{F_c}{A} = \frac{\hbar c \pi ^2}{20 a^4} \left (\frac{-1}{12} \right ) = - \frac{\hbar c \pi ^2}{240 a^4}  
\end{equation} 

\markright{\ptepauthor. \pteptitle}
\section{Shelter Island (1947)}
\markright{\ptepauthor. \pteptitle}

\subsection{The Shelter Island conference}
From June 2 to 4, 1947, the first international physics conference after World War II was held at Shelter Island. The conference brought together 24 physicists from the Manhattan Project, including Bethe, Bohm, Breit, Feynman, Kramers, Lamb, von Neumann, Pauling, Rabi, Schwinger, Teller, Uhlenbeck, Weisskopf and Wheeler. Oppenheimer acted as \linebreak congress master of ceremonies. The participants were received as celebrities, and the conference made a significant impact in the press. Despite high expectations, the conference ended in disappointment.

\begin{figure}[h]
\centering
\includegraphics[scale=0.5]{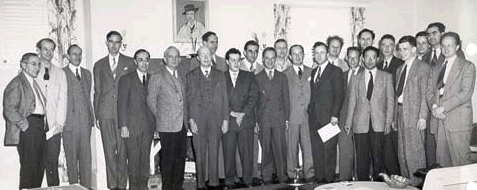}
\vspace{-10pt}
\caption{Shelter Island Conference participants}
\vspace{-5pt}
\end{figure}

Two important experimental measures were presented at the Shelter Island conference: the Lamb shift and the anomalous magnetic moment of the electron. Lamb \cite{Lamb1947} presented an experiment that showed that the 2S1/2 and the 2P1/2 energy levels of the hydrogen atom were not identical; instead they differed by about 1000 MHz. Rabi’s team \cite{Rabi1947} presented a 0.1\% anomaly in the hyperfine structure of hydrogen. Later, Breit \cite{Breit1947} interpreted this anomaly as the anomalous magnetic moment of the electron (g-factor). 

These two measurements contradicted the Dirac equation. Meeting participants assumed that Dirac's theory of the electron was incomplete and proposed that these effects were due to the quantization of the electromagnetic field. It was also assumed that these discrepancies could be calculated using the QFT and that the infinities of this theory could be corrected using renormalization techniques. This was the origin of Quantum Electrodynamics (QED).

If Dirac had been at the Shelter Island conference, the story of the QED would have been very different, as he said: \textbf{\textit{"Renormalization is just a stop-gap procedure. There must be some fundamental change in our ideas, probably a  \linebreak change just as fundamental as the passage from Bohr’s orbit theory to quantum mechanics. When you get a number turning out to be infinite which ought to be finite, you should admit that there is something wrong with your equations, and not hope that you can get a good theory just by doctoring up that number.”}} \cite{DiracI}

\subsection{Bethe's calculation of the Lamb Shift}
On the train trip home after the conference ended, Bethe \linebreak starred in one of the most epic moments in the history of theoretical physics. As recalled by Bethe: \textbf{\textit{"I said to myself, well, let's try to calculate that Lamb shift. And indeed, once the conference was over, I traveled by train to the General Electric research lab. And on the train I figured out how much that difference might be.”}} \cite{Betheyt}

The paper published by Bethe in 1947 \cite{Bethe1947} was a short three pages where he proposed this equation for the Lamb shift. 
\begin{equation}
W_{ns'} = \frac{8}{3\pi}\left(\frac{e^2}{\hbar c}\right)^3 Ry \ \left(\frac{Z^4}{n^3}\right) \ ln \ \frac{K}{<E_n-E_m>_{Av}}
\end{equation}

Where Ry is the ionization energy of the ground state of hydrogen (13.6 eV) and $<E_n-E_m>_{Av}$ is the average excitation energy for the 2s state of hydrogen. 

In this equation, K is a series that diverges to infinity. Bethe decided to apply renormalization by substituting this infinite value for the finite value of the electron’s energy \linebreak ($K = mc^2$). There is no physical justification for making this change, except that when applying that change, the theoretical value was in good agreement with the new experimental value.
\begin{equation}
W_{ns'} = 136 \ ln \ \left( \frac{K}{kp}\right) = 136 \ ln \ \left(\frac{m_ec^2}{17.8 \ Ry} \right)=1040 \ \mbox{Mhz}
\end{equation}

According to the paper, \textbf{\textit{"The average excitation energy has been calculated numerically by Dr. Stehn and Miss Steward.”}}. \cite{Bethe1947} They proposed a value of 17.8 Ry (242 eV), a value that Bethe considered \textbf{\textit{"an amazingly high value."}} \cite{Bethe1947} Informally, $ln(k_p)$ is also known as 'Bethe logarithm'.

As you can see, Bethe's fantastic calculation is based on data that was calculated later, so Bethe could not have known it on his train journey. His calculation included this value that we suspect that was entered ad hoc to match the theoretical value with the experimental value. In the field of physics, this trick is known as a 'fudge factor'.

\subsection{Schwinger's numerology}
A few months after Bethe calculated the value of the Lamb shift, Schwinger devised an even more epic calculation. He published a one-page paper \cite{Schwinger1948} with a simple theoretical value for the electron g-factor, just $\alpha/2\pi$ . 

\begin{equation} 
g = 1 + \frac{\alpha}{2 \pi}
\end{equation}
\begin{equation} 
	\begin{cases}
& g \ (theor) = 1.00116 \\
& g \ (exp) = 1.00119 \\
	\end{cases}
\end{equation}

This value, known as the "Schwinger factor", was in good agreement with the experimental value published by Kush and Foley \cite{Kusch1948}. Schwinger did not explain how he got that value because \textbf{\textit{"a paper dealing with the details of this theory and its applications is in course of preparation."}} \cite{Schwinger1948}

The Schwinger factor had a significant impact on the scientific community due to its simplicity and accuracy. Everyone waited expectantly for the fabulous new theory he had used to calculate this factor. Schwinger’s theory must signify a revolution in modern physics. But, the days passed, and Schwinger did not publish his theory.

Why did not he publish this long-awaited theory? We suspect that Schwinger did not publish the theory because he had no theory. How did he obtain such a spectacular result without a theory? We suspect that he used a technique known as numerology. Schwinger assumed that the g-factor should be directly related to the fine structure constant ($\alpha$), which has an approximate value of 0.7\%. Dividing this value by 6 provides an approximate value of 0.1\%, which is the value obtained by Rabi \cite{Rabi1947}. And $2\pi$ is about 6.

\markright{\ptepauthor. \pteptitle}
\section{Pocono (1948)}
\markright{\ptepauthor. \pteptitle}

\subsection{The Pocono conference}
The Pocono conference took place from March 30 to April 2, 1948. This conference was attended by the same participants as the Shelter Island conference, as well as three of the greatest physicists of the time: Bohr, Dirac and Fermi. As with the Shelter Island conference, the expectations were high due to recent progress from Bethe and Schwinger. And as in the Shelter Island conference, the results were again disappointing.

The conference expectations were focused on \linebreak Schwinger’s presentation. Everyone hoped that he would finally explain the elegant way in which the Schwinger factor had been calculated. 

But Schwinger’s presentation lasted for five unbearable hours and comprised a series of complex, totally incomprehensible equations. Oppenheimer expressed his displeasure: \textbf{\textit{ “others gave talks to show others how to do the calculation, while Schwinger gave talks to show that only he could do it.”}} \cite{Milton} Gradually, the attendees left the presentation until only Bethe and Fermi remained. The overall feeling was one of disappointment, as it was clear that Schwinger’s theory was not based on an elegant solution.

The next day, Feynman presented his theory, explaining for the first time his famous Feynman diagrams. However, the attendees did not respond positively to this presentation. Feynman was convinced of the validity of his calculations simply because they produced the correct results.

Bethe remembers the conference like this: \textbf{\textit{“At Pocono, Schwinger and Feynman, respectively presented their theories. (...) Their theories seemed to be totally different. (...) Schwinger's was closely connected to the known quantum electrodynamics, so Niels Bohr, who was in the audience, immediately was convinced this was correct. And then Feynman came with his completely new ideas, which among other things involved positrons going backwards in time. And Niels Bohr was shocked, that couldn't possibly be true, and gave Feynman a very hard time.”}} \cite{Mehra}

Feynman’s recollection of the conference is also enlightening: \textbf{\textit{“This meeting at Pocono was very exciting, because Schwinger was going to tell how he did things and I was to explain mine. (...). We could talk back and forth, without going into details, but nobody there understood either of us. (...) When he tried to explain his theory, he encountered great difficulty. (...) As soon as he would try to explain the ideas physically, the wolves would descend on him, he had great difficulty. Also, people were getting more and more tired (...) I didn't have a mathematical scheme to talk about. Actually I had discovered one mathematical expression, from which all my diagrams, rules and formulas would come out. The only way I knew that one of my formulas worked was when I got the right result from it. (...) I said in my talk: "This is my mathematical formula, and I'll show you that it produces all the results of quantum electrodynamics." immediately I was asked: "Where does the formula come from?' I said, "It doesn't matter where it comes from; it works, it's the right formula!" "How do you know it's the right formula?" "Because it works, it gives the right results!" "How do you know it gives the right answers?" ' (...) They got bored when I tried to go into the details. (...)  Then I tried to go into the physical ideas. I got deeper and deeper into difficulties, everything chaotic. I tried to explain the tricks I had employed. (...)  I had discovered from empirical rules that if you don't pay attention to it, you get the right answers anyway, and if you do pay attention to it then you have to worry about this and that.”}} \cite{Mehra2}

After the disappointing explanations of Schwinger and Feynman, the scientists returned home, aware of the need for a new unified QED theory that could elegantly explain Bethe’s Lamb shift results and the Schwinger factor for the anomalous magnetic moment of the electron. 

Upon his return to Princeton, Oppenheimer received a paper from a Japanese physicist named Tomonaga with a third QED theory. Now, there were three QED theories, and all of them were inconsistent and incompatible with one other.

\subsection{Dyson’s series}
After the Pocono meeting, the physics community searched for a unified and covariant QED theory. The person in charge of addressing this problem was a young  English scientist \linebreak named Dyson. He managed to reconcile the three QED \linebreak theories in his paper “The Radiation Theories of\linebreak Tomonaga, Schwinger, and Feynman.” \cite{Dyson1949}

Dyson proposed that the Heisenberg S-matrix could be used to calculate the electron’s g-factor, transforming it into a series called the Dyson's series \cite{Dyson19492}. The Dyson's series was an infinite series of powers of alpha, where the first coefficient was precisely the Schwinger factor, and where each coefficient could be calculated by solving a certain number of Feynman diagrams.
\begin{equation}
g = 1 + C_1 \left(\frac{\alpha}{\pi}\right) + 
C_2 \left(\frac{\alpha}{\pi}\right)^2 + 
C_3 \left(\frac{\alpha}{\pi}\right)^3 + 
C_4 \left(\frac{\alpha}{\pi}\right)^4 +
C_5 \left(\frac{\alpha}{\pi}\right)^5 ...
\end{equation}

Dyson's theory based on Feynman diagrams provided the solution his colleagues were waiting for. The enthusiasm returned to the American scientific community.

\subsection{Internal criticism}
However, not all scientists were excited about Feynman’s and Dyson’s results. The primary critic of this new QED theory was Dirac: \textbf{\textit{“How then do they manage with these incorrect equations? These equations lead to infinities when one tries to solve them; these infinities ought not to be there. They remove them artificially. (...) Just because the results happen to be in agreement with observations does not prove that one's theory is correct.”}} \cite{Dirac1}

Another critic was Oppenheimer, as Dyson relates: \linebreak \textbf{\textit{“When after some weeks I had a chance to talk to Oppenheimer, I was astonished to discover that his reasons for being uninterested in my work were quite the opposite of what I had imagined. I had expected that he would disparage my program as merely unoriginal, a minor adumbration of\linebreak Schwinger and Feynman. On the contrary, he considered it to be fundamentally on the wrong track. He thought adumbrating Schwinger and Feynman to be a wasted effort, because he did not believe that the ideas of Schwinger and Feynman had much to do with reality. I had known that he had never appreciated Feynman, but it came as a shock to hear him now violently opposing Schwinger, his own student, whose work he had acclaimed so enthusiastically six months earlier. He had somehow become convinced during his stay in Europe that physics was in need of radically new ideas, that this quantum electrodynamics of Schwinger and Feynman was just another misguided attempt to patch up old ideas with fancy mathematics.”}} \cite{DysonOpp}

According to Dyson, Fermi also did not agree with this new way of conducting science: \textbf{\textit{“When Dyson met Fermi, he quickly put aside the graphs he was being shown indicating agreement between theory and experiment. His verdict, as Dyson remembered, was “There are two ways of doing calculations in theoretical physics. One way, and this is the way I prefer, is to have a clear physical picture of the process you are calculating. The other way is to have a precise and self-consistent mathematical formalism. You have neither.” When a stunned Dyson tried to counter by emphasizing the agreement between experiment and the calculations, Fermi asked him how many free parameters he had used to obtain the fit. Smiling after being told “Four,” Fermi remarked, “I remember my old friend Johnny von Neumann used to say, with four parameters I can fit an elephant, and with five I can make him wiggle his trunk.” There was little to add.”}}
\cite{Fermi}

Feynman’s response to these criticisms is well known:
 \textbf{\textit{“Shut up and Calculate!”}} \cite{Feynmanbook}

\markright{\ptepauthor. \pteptitle}
\section{Oldstone (1949)}
\markright{\ptepauthor. \pteptitle}

\subsection{The Oldstone conference}
From April 11 to 14, 1949, a third conference was held at Oldstone, with the same participants as the Shelter Island and Pocono conferences. As on the previous occasions, the Oldstone conference began with great expectations, this time based on Dyson’s advances. As with the previous conferences, the results were disappointing.

The star of the Oldstone conference was Feynman, who used his immense charisma to present Dyson’s theory as the definitive formalism of the QED theory. From that moment on, Feynman’s diagrams became a popular tool among American physicists, and Feynman became the leader of this new generation of scientists.

In parallel to the QED consolidation, the conference presented important experimental results on subatomic particles that were called pi-mesons or pions. These particles had been discovered thanks to the new synchrocyclotron particle accelerator at the University of Berkeley. Interest in QED rapidly declined due to its extreme complexity and lack of practical utility, while the pions became the primary focus. As a result, Oppenheimer decided not to convene any further QED conferences; instead, he created the International Conference of High Energy Physics (ICHEP).

\subsection{Feynman diagrams in QCD}
New research in high energy physics resulted in Quantum Chromodynamics (QCD), the Electroweak Theory (EWT) \linebreak and the Standard Model of particle physics. All these developments relied heavily on the use of Feynman diagrams. However, the Feynman diagrams are only valid when the coupling constant has a very low value. Obviously if $\alpha > 1$, the Dyson's series would diverge. 

But, in the case of fermions, the coupling constant is \linebreak greater than one, so the Dyson's series would diverge. This means that it is not mathematically legitimate to use Feynman diagrams for these calculations.

In 1951, Feynman himself warned Fermi of this problem: \textbf{\textit{”Don’t believe any calculation in meson theory that uses a Feynman diagram.”\cite{Feynletter}}}

\subsection{Improved Lamb shift calculation}
In 1950, Bethe \cite{Bethe1950} improved the "Bethe logarithm" from 17.8 Ry to 16.646 Ry and incorporated the new factors of Kroll, Feynman, French and Weisskopf into the Lamb shift calculation. With these changes, he obtained an improved value of 1052 Mhz.   
\begin{equation}
W_{ns'} = \frac{\alpha^3}{3\pi \ Ry} \left( \ ln \ \frac{mc^2}{16.646 \ Ry} - ln \ 2 + \frac{5}{6} - \frac{1}{5} \right)
\end{equation}

It was assumed that the difference between the theoretical and experimental values of the Lamb shift was due to relativistic corrections. These corrections were calculated by Baranger in 1951  \cite{Baranger1951} 
\begin{equation}
\Delta W = \alpha ^ 4 Ry \left(1+ \frac{11}{128} - \frac{ln \ 2}{2}\right) = 6.894 \ \mbox{Mhz}
\end{equation}
Baranger obtained a theoretical Lamb shift value of 1058.3 MHz, in good agreement with the new  the experimental value of $1061 \pm 2 $ Mhz.

\markright{\ptepauthor. \pteptitle}
\section{Fourth-order correction (1950-1957)}
\markright{\ptepauthor. \pteptitle}

\subsection{The Kroll \& Karplus calculation}
In 1949, Gardner and Purcell obtained a new experimental result for the g factor of 1.001,146 \cite{Gardner1949}. With this new experimental value, the Schwinger factor was no longer considered accurate. Feynman seized this new crisis as an opportunity to demonstrate the validity of Dyson's reformulation of the QED, where the renormalization of infinities could be performed consistently. 

In 1950, Karplus and Kroll \cite{Karplus1950} completed these complex calculations and published a value of -2,973 for the second coefficient in the Dyson series.

\begin{equation} 
	\begin{cases}
& g =  1 + \dfrac {\alpha}{2 \pi} - 2.973 \left(\dfrac {\alpha}{\pi}\right)^2  \\
& g \ (theor) =  1.001,147 \\
& g \ (exp) = 1.001,146 \\
	\end{cases}
\end{equation}

For the second time, the new theoretical value was in good agreement with the new experimental value.

As indicated in the paper, \textbf{\textit{“The details of two independent calculations which were performed so as to provide some check of the final result are available from the authors.”}} \cite{Karplus1950} The calculations had been performed independently by two teams of mathematicians who had obtained the same result. Therefore, it was impossible that there were any errors in the calculations. Nor was it possible to imagine that a theoretical result that was identical to the experimental result could have been achieved by chance.

This was the definitive test. QED had triumphed. Feynman’s prestige dramatically increased and he began to be \linebreak mentioned as a candidate for the Nobel Prize.

\subsection{Dyson's divergence}
In 1952, two years after this great success, Dyson published a paper entitled “Divergence of Perturbation Theory in \linebreak Quantum Electrodynamics” where he states:\textbf{\textit{ “An argument is presented which leads tentatively to the conclusion that all the power-series expansions currently in use in quantum electrodynamics are divergent after the renormalization of mass and charge.”}} \cite{Dyson1952}

The creator of the QED theory stated that his Dyson's series was divergent. As Dyson admitted: \textbf{\textit{"That was of course a terrible blow to all my hopes. I really meant that this whole program made no sense."}} \cite{Schweber} After the publication of this paper, Dyson moved to England, abandoned this line of research and dedicated the rest of his career to other areas of physics. 

Surprisingly, Dyson’s claim that the series was divergent did not diminish QED’s credibility. 

\subsection{The Witch-Hunt}
 In 1949, the USSR had obtained the atomic bomb thanks to information provided by Fuchs, a Manhattan Project researcher with communist sympathies. Senator McCarthy began a witch hunt where espionage accusations became \linebreak widespread among the American scientific community. \linebreak Oppenheimer was accused of espionage and treason. He was tried and found not guilty, but his position as chief advisor at the AEC was removed.
 
 Now the scientists’ own lives were in jeopardy. Definitely, 1952 was a bad year to admit the complete failure of the main theoretical research program in modern physics.

 The witch hunt ended in 1957, when the Russians sent the Sputnik satellite into space and the US government realized that it needed scientists to create NASA and win the space race.

\subsection{The infamous paper}
In 1956, Franken and Liebes \cite{Franken1956} published new, more precise experimental data that provided a very different g-factor value (1.001,165). This value was higher than the Schwinger factor, so the value of the second coefficient that had been calculated by Kroll and Karplus not only did not improve the Schwinger factor; it made the calculation worse.

With the new experimental data, the value of the second coefficient of the series should have been +0.7 instead of \linebreak -2.973. The difference between these values was huge and unjustifiable. The probative force of QED was upended. In addition, there was no explanation for why Kroll and Karplus’s calculation provided the exact expected experimental value when that value was incorrect. It was evident that the QED calculations had matched the experimental data because they were manipulated. It was a fraud, a scandal.

Karplus and Kroll confessed that they had not independently reached the same result; instead, they had reached a consensus result. Therefore, it was possible that there were errors in the calculation. 

According to Kroll: \textbf{\textit{“Karplus and I carried out the first major application of that program, to calculate the fourth order magnetic moment, which calculation subsequently \linebreak turned out to have some errors in it, which has been a perpetual source of embarrassment to me, but nevertheless the paper I believe was quite influential. (...) The errors were arithmetic (...) We had some internal checks but not nearly enough. (...) it was refereed and published and was a famous paper and now it’s an infamous paper.”}}\cite{Kroll}

Feynman's version of these events does not fully correspond to reality:\textbf{\textit{ “It took two ‘independent’ groups of physicists two years to calculate this next term, and then another year to find out there was a mistake - experimenters had measured the value to be slightly different, and it looked for a while that the theory didn't agree with experiment for the first time, but no: it was a mistake in arithmetic. How could two groups make the same mistake? It turns out that near the end of the calculation the two groups compared notes and ironed out the differences between their calculations, so they were not really independent”.}} \cite{Feynmanbook}

\subsection{Petermann's solution}
Petermann \cite{Petermann1957} detected an error in the Kroll and Karplus calculations (one that no one had detected in the seven years since the article was published). He made the correct calculation and obtained a result of -0.328, which was almost ten times lower than the previous calculation. 

\begin{equation} 
	\begin{cases}
& g =  1 + \dfrac {\alpha}{2 \pi} - 0.328 \left(\dfrac {\alpha}{\pi}\right)^2  \\
& g \ (theor) =  1.001,159,6 \\
& g \ (exp) = 1.001,165 \\
	\end{cases}
\end{equation}

The same error was independently detected by Sommerfield \cite{Sommerfield1957}. Once again, two independent calculations provided the same theoretical value. 

For the third time, the new theoretical value was in good agreement with the new experimental value.

\markright{\ptepauthor. \pteptitle}
\section{The Unpublished Feynman diagram IIc}
\markright{\ptepauthor. \pteptitle}

\subsection{Karplus and Kroll's paper}

At this point, we have doubts about everything that was reported, so we reviewed the original article published by Kroll and Karplus as well as the corrections of Petermann and Sommerfield published seven years after.

\begin{figure}[h]
\centering
\includegraphics[scale=0.50]{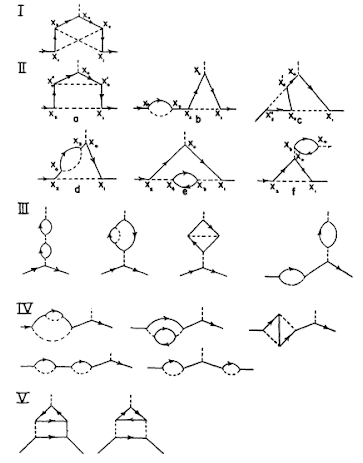}
\vspace{-10pt}
\caption{Feynman diagrams for the fourth-order corrections}
\label{fig:feynman}
\vspace{-5pt}
\end{figure}

Karplus and Kroll's paper consists of 14 pages full of complex mathematical calculations\cite{Karplus1950}. On the second page of the document, the authors indicate that to obtain the coefficient it is necessary to calculate 18 Feynman diagrams grouped in five groups (I, II, III, IV and V). However, on pages 3 and 4, they argue that groups III, IV and V are not necessary. 

Therefore, it is only necessary to calculate seven Feynman diagrams, identified as I, IIa, IIb, IIc, IId, IIe, IIf. A lot of calculations are done between pages 4 and 11 that only serve to show that diagrams IIb and IIf are not necessary either. Therefore, it is only necessary to calculate five Feynman diagrams (I, IIa, IIc, IId, IIe).

The calculation of diagrams IIe (0.016) and IId (-0.090) are performed on pages 11 and 12 respectively. It follows that: \textbf{\textit{“The expressions for I, IIa and IIc become successively more complicated and very much more tedious to \linebreak evaluate and cannot be given in detail here.”}} \cite{Karplus1950} In other words, the complete calculation of three of the five diagrams was never published. On page 13, the results of the three remaining diagrams are shown (I = -0.499, IIa = 0.778 and IIc = -3.178). Finally, page 14 of the paper presents the "Summary of Results" with the results of each of the five diagrams.

\begin{equation}
C_2 = I + IIa + IIc + IId + IIe = -2,973
\end{equation}
\vspace{-20pt}
\renewcommand{\arraystretch}{1.3}
\begin{table}[h]
\centering
\begin{tabular}{ |ccccc|c|} 
\hline
I & IIa & IIc & IId & IIe & Total \\
\hline
-0.499 & 0.778 & -3.178 & -0.090 & 0.016 & -2.973 \\
\hline
\end{tabular}
\vspace{-5pt}
\caption{\label{tab:KK} Values of the five Feynman diagrams.}
\end{table}

Diagrams IId and IIe were the only diagrams whose calculations are included in the document; however, its values were completely irrelevant. Diagrams I and IIa practically cancel each other out. IIc was the dominant diagram and consisted of four components:
\begin{equation} \label{eq:KK}
II_c = -\frac{323}{24} + \frac{31}{9} \pi ^2 -  \frac{49}{6} \pi ^2 ln (2) +  \frac{107}{4} \zeta(3) = -3,178
\end{equation}
\vspace{-12pt}
\renewcommand{\arraystretch}{1.3}
\begin{table}[h]
\centering
\begin{tabular}{ |cccc|c|} 
\hline
Constant & $\pi^2$ & $\pi^2ln2$ & $\zeta(3)$ & Total  \\
\hline
-13,458 & 33,995 & -55,868 & 32,153 & -3,178 \\
\hline
\end{tabular}
\vspace{-5pt}
\caption{\label{tab:KKIIc} Value of the four components of Feynman diagram IIc.}
\end{table}

The four components of IIc have abnormally high values \newline (-13, 34, -55 and 32) which surprisingly compensate for each other, resulting in -3,178, an order of magnitude lower. It is not possible to say anything more about the calculation of diagram IIc because the complete calculation was never published.

\subsection{Petermann's numerical calculation}
Petermann was the first person to identify an error in the original calculation of Karplus and Kroll. He performed a numerical analysis of the five Feynman diagrams. He found that the solution of diagram IIc was clearly wrong, since its value was outside the limits. The rest of the diagrams were within limits: \textbf{\textit{“The numerical results for the terms I, IIa, IIc, IId, IIe in the work by Karplus and Kroll have been checked by rigorous upper and lower bounds. Whereas every other term fell well between these bounds, agreement could not be obtained for diagram IIc. (...) The numerical value for this term has been found to satisfy IIc = -1.02 +/- 0.53.”}} \cite{Peter1}

Petermann published a second paper where he adjusted his calculations: \textbf{\textit{”the diagram IIc is found to satisfy IIc = -0.60 +/- 0.11 in contradiction with the value -3.18 given by the previous authors.”}} \cite{Peter2}

Between the publication of these two papers, Petermann communicated privately to Sommerfield the result of another calculation: \textbf{\textit{"Petermann has placed upper and lower \linebreak bounds on the separate terms of Karplus and Kroll. He finds that their value for IIc does not lie within the appropriate bounds. Assuming the other terms to be correct, he concludes that the result is -0.53 +/- 0.37.”}} \cite{Sommerfield1957}

Petermann worked for three months following a numerical methodology that allowed him to narrow the margin of error in diagram IIc. Surprisingly, fourteen days after his third numerical calculation, he made an unexpected change in his methodology and published the exact analytical calculation, with no margins of error.

The articles published by Petermann on the calculation of the Feynman diagram IIc are summarized in the following table:

\renewcommand{\arraystretch}{1.3}
\begin{table}[ht]
\begin{tabular}{|c|c|c|p{2.7cm}|}
\hline
Date & IIc & Method & Publication \\
\hline
28/5 & -1.02 +/- 0.53 & Numerical & Nuclear Phys. 3 \\
\hline
1/7 & - 0.53 +/- 0.37 & Numerical & Phys. Rev. 107,  Note added in proof. \\  
\hline
3/8 & -0.60 +/- 0.11 & Numerical & Nuclear Phys. 5 \\
\hline
17/8 & -0,564 & Analytical & Helvetica Physica Acta 30 \\
\hline
\end{tabular}
\vspace{-10pt}
\caption{\label{tab:Petpapers} Petermann's publications.}
\vspace{-10pt}
\end{table}

\subsection{Sommerfield and the Green’s functions}

After the publication of the new experimental value by \linebreak Franken and Liebes \cite{Franken1956}, Schwinger commissioned a \linebreak 22-year-old student named Sommerfield to redo the Kroll and Karplus calculations. Schwinger proposed using his own \linebreak method based on Green's functions instead of using Feynman diagrams. According to Sommerfield: \textbf{\textit{"Julian assigned us three problems, one of which involved the anomalous magnetic moment (...). At my meeting with him, he suggested that I continue the calculation of the anomalous magnetic moment to the next fourth order (...). Schwinger wanted me to use the other method, while respecting gauge invariance at every step. Many years later Roy Glauber told me that the faculty was not entirely happy that a graduate student had been given such a problem."}} \cite{Sommer100}

In May 1957, Sommerfield sent a two-page paper to the Physical Review Journal where he published his results:
\begin{equation}
C_2 = \frac{197}{144} + \frac{1}{12} \pi ^2 - \frac{1}{2} \pi ^2 ln (2) + \frac{3}{4} \zeta(3) = -0,328
\end{equation}

Sommerfield's paper does not include the calculations \linebreak performed, but the author states that: \textbf{\textit{“The present calculation has been checked several times and all of the auxiliary integrals have been done in at least two different ways.”}} \cite{Sommerfield1957} As a guarantee that the calculations were correct. 

He also states that: \textbf{\textit{“The discrepancy has been traced to the term I y IIc of Karplus and Kroll.”}} \cite{Sommerfield1957} But this statement about the origin of the error cannot be deduced from Sommerfield's calculations, since he used Green's functions instead of Feynman diagrams. 

In 1958 Sommerfield published the full calculation of the g-factor calculations in an extensive 32-page paper as part of his doctoral thesis \cite{Sommer2}. He used Green's functions instead of Feynman diagrams. Then, the calculation of the enigmatic Feynman diagram IIc does not appear in his paper.

As Schwinger states: \textbf{\textit{“Interestingly enough, although \linebreak Feynman-Dyson methods were applied early [by Karplus and Kroll], the first correct higher order calculation was done by Sommerfield using [my] methods.”}}.\cite{Mehra} In the third volume of "Particles, Sources, and Fields" published in 1989 \cite{Schwinger1986}, Schwinger devoted more than 60 pages to a detailed calculation of the second coefficient of Dyson series getting exactly the same result. But, once again, using Green's functions instead of Feynman diagrams.

\subsection{Petermann's definitive correction}

The definitive solution to the problem was presented in 1957 by Petermann in a paper published in the Swiss journal Helvetica Physica Acta \cite{Petermann1957}. 

The article was signed by a single author due to an internal conflict between the researchers. As Sommerfied recalls: \textbf{\textit{"In the meantime Schwingerian Paul Martin had gone to the Niels Bohr Institute in Copenhagen and had spoken to Andre Petermann, a postdoc with the Swedish theoretician Gunnar Kallen. Martin told Petermann about my work (...)  In the end, however, after both of our calculations were completely finished they were in agreement with each other but not with Karplus and Kroll. We agreed to cite each other's work when published. However, Schwinger and Kallen had had a somewhat acrimonious discussion (...) and Kallen had forbidden Petermann to mention my work. Petermann's apology to me was profuse."}}. \cite{Sommer100} Although the paper was signed by a single author, Petermann acknowledges that the result was obtained by consensus: \textbf{\textit{“The new fourth order correction given here is in agreement with: (a) The upper and lower bounds given by the author. (b) A calculation using a different method, performed by C. Sommerfield. (c) A recalculation done by N. M. Kroll and collaborators.”}}. \cite{Petermann1957} Petermann's final result was identical to the \linebreak Sommerfield's result published three months earlier.

\begin{equation}
C_2 = I + IIa + IIc + IId + IIe =  -0,328
\end{equation}
\vspace{-20pt}
\renewcommand{\arraystretch}{1.3}
\begin{table}[ht]
\centering
\begin{tabular}{ |ccccc|c|} 
\hline
I & IIa & IIc & IId & IIe & Total \\
\hline
-0.467 & 0.778 & -0.564 & -0.090 & 0.016 & -0.328 \\
\hline
\end{tabular}
\vspace{-5pt}
\caption{\label{tab:Pet} Corrected values of the five Feynman diagrams.}
\end{table}
\begin{equation}  \label{eq:PS}
II_c = -\frac{67}{24} +  \frac{1}{18} \pi ^2 +  \frac{1}{3} \pi ^2 ln (2) -  \frac{1}{2} \zeta(3) =  -0,564 
\end{equation}
The following table compares the calculations of the four components of the Feynman IIc diagram made by Karplus and Kroll (Equation \ref{eq:KK}) with the calculations made by Petermann (Equation \ref{eq:PS}).

\renewcommand{\arraystretch}{1.3}
\begin{table}[ht]
\centering
\begin{tabular}{ |c|cccc|c|} 
\hline
&  Const. & $\pi^2$ & $\pi^2 \ ln(2)$ & $\zeta(3)$ & Total \\ 
\hline
K\&K & -13,458 & 33,995 & -55,868 & 32,153 & -3,178 \\
\hline
P\&S & -2,791 & 0,548 & 2,280 & -0,601 & -0,564 \\
\hline
Diff. & 10,667 & -33,447 & 58,148 & -32,754 & 2,614 \\
\hline
\end{tabular}
\vspace{-10pt}
\caption{\label{tab:compIIc} Comparative components of Feynman diagram IIc.}
\end{table}

The corrections are huge, one or two orders of magnitude for each component of diagram IIc. We cannot know the origin of these discrepancies because the correction calculations were also not published.

\subsection{Smrz \& Uleha}
In 1960, Smrz \& Uleha published a short paper of two pages where the situation generated in 1957 by Petermann's correction is explained. The authors states that they performed an independent calculation of the Feynman IIc diagram and obtained exactly the same result as Petermann. \textbf{\textit{“Since the considerable difference between the original value of the magnetic moment (Karplus \& Kroll \cite{Karplus1950}) and the values calculated later (Petermann \cite{Petermann1957}) originates in the calculation of the contribution from the third diagram, only the value of this contribution was determined by the standard technique and the above regularization in the infra-red region. The contribution from the third diagram (-0.564) is in complete agreement with Petermann's value.”}}  \cite{Smrz} 

Unfortunately when looking for the reference of the work with the calculations, it has not been published either:
\textbf{\textit{"Smrz P.: Diploma thesis, Faculty of Tech. and Nucl. Physics, Prague 1960, unpublished."}}  \cite{Smrz} 

\subsection{Terentiev}
In 1962, Terentiev published a long paper of 50 pages \cite{Terentiev}. The paper is only in Russian and there is no English translation. We identify the equation 60 of the paper as the second coefficient of the Dyson's series, with the same expression and value obtained by Petermann. Analyzing the document, we interpret that this equation is the result of the sum of nine other equations identified as equations 22, 24, 27, 31, 33, 47, 51, 58 and 59. There are nine equations instead of the five Feynman diagrams of Karplus and Kroll and none of the these equations correspond to the Feynman Diagram IIc. 

\subsection{Barbieri \& Remiddi}
In 1972, Barberi \& Remeddi published a long 93-page paper where they performed a recalculation of the Feynman diagrams corresponding to the fourth-order coefficient and confirmed the results obtained by Petermann. The authors give the descriptive name of "Corner Graphs" to the the Feynman IIc. On the specific calculations, the authors state the following: \textbf{\textit{"Integrations by parts, differentiations and so on, was done by computer, using the program SCHOONSCHIP of Veltman.”}} \cite{Remiddi} That is, they used a computer program to perform the mathematical calculations, but they did not publish the source code used, so, again, it is not possible to replicate the calculations.

On the first page of Barbieri \& Remiddi paper there is a reference to Terentiev's paper, where the authors claim that Terentiev's results were incorrect and manipulated: \textbf{\textit{“Dispersion relations are used in the Terentiev work only to write down suitable multiple integral representations, which are in general manipulated to get the final result, without explicitly evaluating the discontinuities. The problem of infra-red divergences has been further overlooked, and many of the intermediate results are wrong, even if somewhat ad hoc compensations make the final result correct.”}}  \cite{Remiddi}

\subsection{Summary of the situation}

The history of this calculation is surrounded by  big errors and inexplicable coincidences.

\begin{itemize}

\item The original calculation of the Feynman diagram  IIc published in 1950 was completely wrong. 

\item Karplus and Kroll stated that the calculation had been performed by two teams independently. This statement was made to give guarantees about the validity of the calculations, and yet it turned out to be false.

\item Despite having published a completely wrong result, the prestige of Karplus and Kroll was not affected at all. On the contrary, both enjoyed brilliant careers full of awards and recognition for their professional achievements.

\item Karplus and Kroll's miscalculation was consistent with the experimental value previously published by \linebreak Gardner and Purcell, even though that experimental \linebreak value was also wrong.

\item The error in the calculation was not reported until seven years after its publication. 

\item The error in the calculation was detected just when a new experimental value was published by Franken and Liebes. The corrected theoretical value also coincided with the new experimental value.

\item Neither the original calculation of the Feynman diagram IIc nor its subsequent correction has been published to date.

\item The recalculations of the Feynman diagram IIc of Smrz \& Uleha and Barbieri \& Remiddi have not been published either.

\item The recalculation of Terentiev were revisited ten years later by Remmidi, who claims that these calculations were wrong and manipulated with ad-hoc compensations to make the final result correct.
\end{itemize}

\renewcommand{\arraystretch}{1.3}
\begin{table}[ht]
\centering
\begin{tabular}{|c|c|p{4.3cm}|}
\hline
Year & Author &  Status of calculations\\
\hline
1950 & Karplus \& Kroll & Wrong and Unpublished \\
\hline
1957 & Petermann & Right but Unpublished \\  
\hline
1957 & Sommerfield & Right but Unpublished \\
\hline
1958 & Sommerfield & Right but using Green's Functions instead of \linebreak Feynman Diagrams \\
\hline
1960 & Smrz \& Uleha & Right but Unpublished \\
\hline
1962 & Terentiev & Wrong intermediate results with ad hoc compensations to make the final result correct \\
\hline
1972 & Remiddi & Right but Unpublished \linebreak Computer calculation  \\
\hline
1989 & Schwinger & Right but using Green's \linebreak Functions \\
\hline
\end{tabular}
\vspace{-10pt}
\caption{\label{tab:Calculations} fourth-order coefficient calculation.}
\vspace{-10pt}
\end{table}
\markright{\ptepauthor. \pteptitle}
\section{The Nobel Prize (1965)}
\markright{\ptepauthor. \pteptitle}

\subsection{Direct g-factor measurement}
In 1953, a research team from the University of Michigan \cite{Louisell1954} proposed a new type of experiment to calculate the magnetic moment of the electron directly from the precession of the free electron spin. This new technique provided more precise experimental values than the previous techniques that were based on atomic levels. 

In 1961, Schupp, Pidd and Crane carried on the experiment and published a new experimental value of 1.0011609. The experiment was revolutionary because of the measured precision, however, the authors were cautious with their results, presenting large margins of error. The explanation for this strange decision is found in the paper: \textbf{\textit{“In deciding upon a single value for a to give as the result of the experiment, our judgement is that we should recognize the trend of the points (...). The value a=0.0011609, obtained in this way, may be compared with a simple weighted average of the data of Table IV, which is 0.0011627. We adopt the value 0.0011609 but assign a standard error which is great \linebreak enough to include the weighted average of Table IV, namely $\boldsymbol{\pm 0.0000020}$. Finally, we combine with this the estimated systematic standard errors (...). This results in a final value of $\boldsymbol{0.0011609 \pm 0.0000024}$”}}. \cite{Schupp1961} 

\begin{figure}[h]
\centering
\includegraphics[scale=0.4]{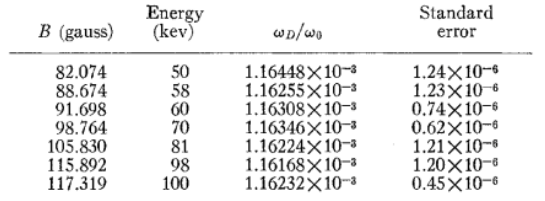}
\vspace{-10pt}
\caption{Table IV, the g-factor anomaly calculated for the various electron energies}
\end{figure}

According to this explanation, the estimated systematic standard error was 0.0000004. If this error had been published, the result would have been $1.0011609 \pm 0.0000004$, leaving Petermann’s theoretical value outside the margin of error and creating a new crisis in the development of QED. 

The authors proposed another possible approach: they averaged the measurements in Table IV, generating a result of $1.0011627 \pm 0.0000024$. But this alternative result also left out of the margin of error the Petermann's theoretical value.

Finally, the authors published a meaningless result. Although they published what they considered to be the correct result (1.0011609), they added a margin of error of \linebreak +0.0000024 to include the average of the actual results. They also added a negative symmetrical margin of error of \linebreak -0.0000024, without any logical; this was the only way to keep Petermann’s theoretical value within the margin of error.

\subsection{The experimenter's bias}
At that time the situation was dramatic again. Predictably, subsequent experiments would discredit the g-factor theoretical value. And after the Kroll and Karplus scandal, the theoretical calculations could not be modified again to adapt them to the experimental data without completely distorting the QED. 

And the moment come in 1963, Wilkinson and Crane published a improved version of the experiment. In the report of the results all the previous cautionary language disappeared. The accuracy of this result was presented as 100 times higher than that of the previous experiment, and the tone of the paper was blunt: \textbf{\textit{“mainly for experimental reasons, we here conclude the 10-year effort of the laboratory on the g factor of the free negative electron.”}} \cite{Wilkinson1963}

Just when QED seemed doomed to disaster, the miracle happened again. This time, the new experimental value was $1,001,159,622 \pm 0.000,000,027$, nearly the same as Petermann’s theoretical value (1,001,159,615).

\begin{figure}[ht]
\centering
\includegraphics[scale=0.5]{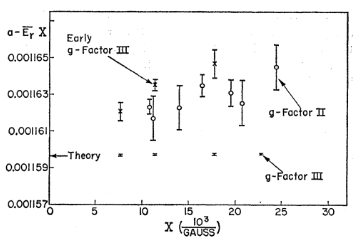}
\vspace{-10pt}
\caption{Experimental values}
\end{figure}

This experimental result is incredibly suspicious. It was obtained after a simple improvement of the previous experiment, and it was conducted at the same University, with the same team, only two years later. It is extremely strange that all the measurements from the previous experiment were outside the range of the new experimental value. Even stranger, the theoretical value fit perfectly within the experimental \linebreak value. Most disturbing, this value is not correct, as was \linebreak demonstrated in later experiments.  

\subsection{The Nobel Prize}

In 1965, Drell and Pagels \cite{Drell} published the first calculation of the third coefficient of the Dyson's series (sixth order correction), that implied solving 72 Feynman diagrams. The result was 0.15, which allowed to slightly improve the theoretical value of the electron g-factor.

\begin{equation} 
	\begin{cases}
& g =  1 + \dfrac {\alpha}{2 \pi} - 0.328 \left(\dfrac {\alpha}{\pi}\right)^2 + 0.15 \left(\dfrac {\alpha}{\pi}\right)^3  \\
& g \ (theor) =  1,001,159,617 \\
& g \ (exp) = 1,001,159,622 \\
	\end{cases}
\end{equation}

For the fourth time, the new theoretical value was in good agreement with the new experimental value. At this point all doubts about QED were cleared and Feynman, Schwinger and Tomonaga were awarded the Nobel Prize in physics in 1965.

\markright{\ptepauthor. \pteptitle}
\section{Sixth-order Correction (1968-1972)}
\markright{\ptepauthor. \pteptitle}

\subsection{Wesley and Rich, preliminary}
In 1968, Rich \cite{Rich1} reevaluated the Wilkinson's experiment and obtained a different result. Other researchers found more errors in the experiment which allowed different interpretations of the results. It became clear that the Wilkinson's experiment had to be repeated to fix these discrepancies. 

In 1970, Rich and Wesley \cite{Rich} repeated Wilkinson's experiment, fixing the detected ambiguities and obtaining a \linebreak result higher that the previous one.

The same year, Brodky \& Drell \cite{Drell2}  recalculated the \linebreak sixth-order factor and obtained a value of +0.55, three times higher that the previous value.

\begin{equation} 
	\begin{cases}
& g =  1 + \dfrac {\alpha}{2 \pi} - 0.328 \left(\dfrac {\alpha}{\pi}\right)^2 + 0.55 \left(\dfrac {\alpha}{\pi}\right)^3  \\
& g \ (theor) =  1,001,159,644 \\
& g \ (exp) = 1,001,159,644 \\
	\end{cases}
\end{equation}

For the fifth time, the new theoretical value was in good agreement with the new experimental value. 

Brodky and Drell summarized the situation: \textbf{\textit{“Quantum electrodynamics has never been more successful in its confrontation with experiment than it is now. There is really no outstanding discrepancy despite our pursuing the limits of the theory to higher accuracy (...) however, and despite its phenomenal success, the fundamental problems of renormalization in local field theory and the nature of the exact solutions of quantum electrodynamics are still to be resolved.”}}\cite{Drell2}

Starting in the 1970s, all the mathematical calculations necessary to obtain the coefficients of the Dyson series were performed by computer. No source code for these calculations has been published, so it is not possible to reproduce any calculations independently.

\subsection{Wesley and Rich, definitive}
The results published by Wesley and Rich were preliminary. In 1971 they published the final results of the experiment that turned out to be higher than expected.\cite{Rich2}

The same year, Levine \& Wright \cite{Levine} recalculated the sixth-order factor and obtained a value of +1.49, three times higher that the previous value.
\begin{equation} 
	\begin{cases}
& g =  1 + \dfrac {\alpha}{2 \pi} - 0.328 \left(\dfrac {\alpha}{\pi}\right)^2 + 1.49 \left(\dfrac {\alpha}{\pi}\right)^3  \\
& g \ (theor) =  1,001,159,655 \\
& g \ (exp) = 1,001,159,657 \\
	\end{cases}
\end{equation}

For the sixth time, the new theoretical value was in good agreement with the new experimental value.

Wesley and  Rich summarized the situation: \textbf{\textit{"The agreement between the experimental measurements and QED \linebreak predictions of the electron g-factor at a level of four parts per billion represents five most accurate comparison \linebreak between theory and experiment in physics. In spite of the unsatisfactory nature of the renormalization aspects of the theory, there has been no significant evidence for a breakdown of QED”.}} \cite{Rich3}

A few months later, Kinoshita \& Cvitanovic \cite{Cvitanovic1} published a new calculation for the sixth-order factor, five times more accurate than Levine \& Wright's previous one. However, this new theoretical value (+1.29)  slightly worsened the agreement with the experimental value.

\begin{equation} 
	\begin{cases}
& g =  1 + \dfrac {\alpha}{2 \pi} - 0.328 \left(\dfrac {\alpha}{\pi}\right)^2 + 1.29 \left(\dfrac {\alpha}{\pi}\right)^3  \\
& g \ (theor) =  1,001,159,653 \\
& g \ (exp) = 1,001,159,657 \\
	\end{cases}
\end{equation}

\subsection{Cvitanovic's conjecture}
According to Dyson, the value of the coefficients of the series should grow exponentially, but this did not happen with the coefficients calculated up to that moment. Cvitanovic observed that calculating the coefficients of the Dyson's series produced many more cancellations than expected. In 1977 he proposed a conjecture \cite{Cvitanovic2} according to which the calculation of the coefficients would remain at values close to one. So far, Cvitanovic's conjecture has not been proven nor has Dyson's argument been shown to be wrong.

On the other hand, in 1978, the Nobel laureate 't Hooft presented another argument, completely different from \linebreak Dyson's argument, but with the same consequences. According to' t Hooft: \textbf{\textit{“We understand how to renormalize the theory to any finite order in the perturbation expansion, but it is expected that this expansion will diverge badly, for any value of the coupling constant."}} \cite{Hooft}

\subsection{Landau pole}
The Landau pole (or the Moscow zero, or the Landau ghost) is the momentum scale at which the coupling constant  of a quantum field theory becomes infinite. Such a possibility was pointed out by Landau in 1955. \cite{Landau}

The asymptotic freedom of QCD was discovered in 1973 by Gross and Wilczek, but asymptotic freedom is not applicable to QED. Landau poles appear in theories that are not asymptotically free, such as QED. In these theories, the renormalized coupling constant grows with energy. In a theory purporting to be complete, this could be considered a mathematical inconsistency.

\markright{\ptepauthor. \pteptitle}
\section{Eight-order Correction (1977-2008)}
\markright{\ptepauthor. \pteptitle}

\subsection{The Penning trap}

In 1977, Van Dyck and Dehmelt of the University of Washington used a new technique known as free electron spin resonance. These measurements were based on a device called a Penning trap, which allowed measurements to be obtained from individual electrons. These experiments improved the previous results by three orders of magnitude, and, again, the new result excluded previous theoretical value. \cite{Dyck1977}

To resolve this new discrepancy, the theoretical physicists needed to calculate the the fourth coefficient (eighth-order correction), which involved solving 891 new Feynman diagrams. In 1981, Kinoshita \& Lindquist \cite{Kinoshita1981} published the first calculation of the eight-order coefficient with a value of -0.8.

In 1982, Levine \cite{Levine2} published a new calculation of the sixth-order coefficient with a value of +1.176, lower that the previous one.
\begin{equation} 
	\begin{cases}
& g =  1 + \dfrac {\alpha}{2 \pi} - 0.328 \left(\dfrac {\alpha}{\pi}\right)^2 + 1.176 \left(\dfrac {\alpha}{\pi}\right)^3 - 0.8 \left(\dfrac {\alpha}{\pi}\right)^4  \\
& g \ (theor) = 1.001,159,652,460 \\
& g \ (exp) = 1.001,159,652,410 \\
	\end{cases}
\end{equation}

For the seventh time, the new theoretical value was in good agreement with the new experimental value.

\subsection{Feynman's Book}

In 1981, Van Dyck and Dehmelt \cite{Dyck1981} published a second experimental value, lower that the previous one. 
\begin{equation} 
	\begin{cases}
& g \ (theor) = 1.001,159,652,460 \\
& g \ (exp) = 1.001,159,652,222 \\
	\end{cases}
\end{equation}
Just these numbers were used in 1985 by Feynman in his famous book titled "QED: The Strange Theory of Light and Matter".  On page 7 of the book, Feynman shows this data and says: \textbf{\textit{"At the present time I can proudly say that there is no significant difference between experiment and theory! (...) To give you a feeling for the accuracy of these numbers, it comes out something like this: If you were to measure the distance from Los Angeles to New York to this accuracy, it would be exact to the thickness of a human hair."}} \cite{Feynmanbook}

However, at the time of publication of Feynamn's book there was a real discrepancy between the theoretical and the experimental value that remained for a decade. Van Dyck and Dehmelt \cite{Dyck1984} published another two new experimental values that increased that discrepancy:
\begin{itemize}
\item $[1984]: 1.001,159,652,193$ \cite{Dyck1984}
\item $[1987]: 1.001,159,652,188,4$ \cite{Dyck1987}
\end{itemize}

\subsection{Analytical sixth-order calculation}

In 1995, Kinoshita's team published a new value of the \linebreak eighth-order correction (-1.557), the double of his first calculation. \cite{Kinoshita1995}. And in 1996, Laporta and Remiddi \cite{Laporta1996} published the analytical calculation of sixth-order (+1.181). This definitive value was eight times higher than the initial calculation of Drell (+0.15). 
\begin{equation} 
	\begin{cases}
& g =  1 + \dfrac {\alpha}{2 \pi} - 0.328 \left(\dfrac {\alpha}{\pi}\right)^2 + 1.181 \left(\dfrac {\alpha}{\pi}\right)^3 - 1.557 \left(\dfrac {\alpha}{\pi}\right)^4  \\
& g \ (theor) =  1.001,159,652,201,2 \\
& g \ (exp) = 1.001,159,652,188,4 \\
	\end{cases}
\end{equation}

For the eighth time, the new theoretical value was in good agreement with the new experimental value.

\subsection{Muon's g-factor}
QED was also used to calculate the anomalous magnetic moment of the muon. The muon is an unstable subatomic \linebreak particle with a mean lifetime of $ 2.2\ \mu s$, making high resolution measurements extremely complex. From 1961 to 1976, CERN made the first measurements of the muon g-factor. The following experiment was carried out at Brookhaven National Laboratory (BNL). The experiment was named E821 and the results were published from 1997 to 2001. Unfortunately, the theoretical value did not match the new experimental value.

New factors were added to adjust the theoretical final result. These new factors came from empirical data obtained from the Standard Model of particle physics. The first coefficient was derived from the interaction of the electron with leptons, the second coefficient was derived from the electroweak interaction and the third coefficient was derived from the electron’s interaction with hadrons.
\begin{equation}
g = g \ (QED) +  g \ (noQED)
\end{equation}
\begin{equation}
 g \ (noQED) =  g \ (\mu, \tau) +  g \ (weak) + g \ (hadron) \end{equation}

The inclusion of these new coefficients in the g-factor calculation improved the theoretical muon g-factor value but worsened the theoretical electron g-factor value. This change created a new discrepancy between the theoretical and experimental value of the electron g-factor.

\subsection{Harvard experiment}
In 2006 a team from Harvard University led by Gabrielse improved the experimental results of Van Dyck and Dehmelt by two orders of magnitude. The Harvard University data were not compatible with previous experimental data provided by the University of Washington. These new data also excluded the theoretical value of the g-factor.

\begin{itemize}
\item $[2006]: 1.001,159,652,180,85(76)$ \cite{Gabrielse2006}
\item $[2008]: 1.001,159,652,180,73(28)$ \cite{Gabrielse2008}
\end{itemize}
\vspace{-10pt}
\begin{figure}[h]
\centering
\includegraphics[scale=0.4]{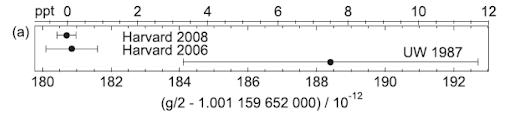}
\vspace{-5pt}
\caption{Harvard vs Washington errors}
\vspace{-5pt}
\end{figure}
\begin{figure}[h]
\centering
\includegraphics[scale=0.35]{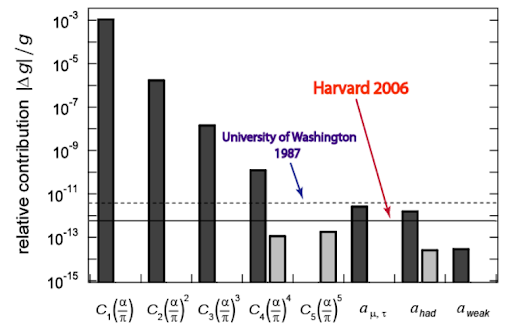}
\vspace{-10pt}
\caption{electron g-factor errors}
\vspace{-10pt}
\end{figure}

In 2007, Kinoshita's team detected an error in his previous calculation of the eighth-order: \textbf{\textit{"Comparing the contributions of individual diagrams of old and new calculations, we found an inconsistency in the old treatment of infrared subtraction terms in two diagrams. Correcting this error leads to the revised value -1.914 for the eight-order term."}} \cite{Kinoshita2007}
\begin{equation} 
	\begin{cases}
& g =  1 + \dfrac {\alpha}{2 \pi} - 0.328 \left(\dfrac {\alpha}{\pi}\right)^2 + 1.181 \left(\dfrac {\alpha}{\pi}\right)^3 \\
& - 1.914 \left(\dfrac {\alpha}{\pi}\right)^4 +  g \ (noQED) \\
& g \ (theor) = 1.001,159,652,182.79 \\
& g \ (exp) = 1.001,159,652,180,73 \\
	\end{cases}
\end{equation}

For the ninth time, the new theoretical value was in good agreement with the new experimental value.

\markright{\ptepauthor. \pteptitle}
\section{Tenth-order correction (1997-2021)}
\markright{\ptepauthor. \pteptitle}

\subsection{Tenth-order correction}

Every new calculations of $g \ (noQED)$ allowed to reduce the discrepancy of the muon g-factor but worsened the discrepancy of the electron g-factor. To resolve this new discrepancy, the theoretical physicists needed to resolve the 12,672 Feynman diagrams of the tenth-order corrections. 

In 2012, Kinoshita's team published the first tenth-order coefficient with a value of +9.16, and in the same paper, they published an improved value of -1.909 for the eighth-order contribution. \cite{Kinoshita2012}
\begin{equation} 
	\begin{cases}
	& g =  1 + \dfrac {\alpha}{2 \pi} - 0.328 \left(\dfrac {\alpha}{\pi}\right)^2 + 1.181 \left(\dfrac {\alpha}{\pi}\right)^3 \\
	&  -1.909 \left(\dfrac {\alpha}{\pi}\right)^4  + 9.16 \left(\dfrac {\alpha}{\pi}\right)^5 	+  g \ (noQED)  \\
& g \ (theor) = 1.001,159,652,181,78 \\
& g \ (exp) = 1.001,159,652,180,73 \\
	\end{cases}
\end{equation}

For the tenth time, the new theoretical value was in good agreement with the new experimental value.

 In 2015, Kinoshita's team published an improved value of the tenth-order coefficient with a value of +7.795, and in the same paper, they published an improved value of -1.912 for the eighth-order contribution. \cite{Kinoshita2015}

\begin{equation} 
	\begin{cases}
	& g =  1 + \dfrac {\alpha}{2 \pi} - 0.328 \left(\dfrac {\alpha}{\pi}\right)^2 + 1.181 \left(\dfrac {\alpha}{\pi}\right)^3 \\
	& -1.912 \left(\dfrac {\alpha}{\pi}\right)^4  + 7.795 \left(\dfrac {\alpha}{\pi}\right)^5 	+  g \ (noQED)  \\
& g \ (theor) = 1.001,159,652,181,64 \\
& g \ (exp) = 1.001,159,652,180,73 \\
	\end{cases}
\end{equation}

For the eleventh time, the new theoretical value was in good agreement with the new experimental value.

In 2017, Laporta \cite{Laporta1100} published his final calculation of the eighth-order coefficient of the Dyson's series with a value of -1.912, the double of the initial estimation (-0.8). The published value had an unnecessary accuracy of 1100 digits, as a proof that this numerical calculation can be considered the definitive result.  

\begin{figure}[h]
\centering
\includegraphics[scale=0.35]{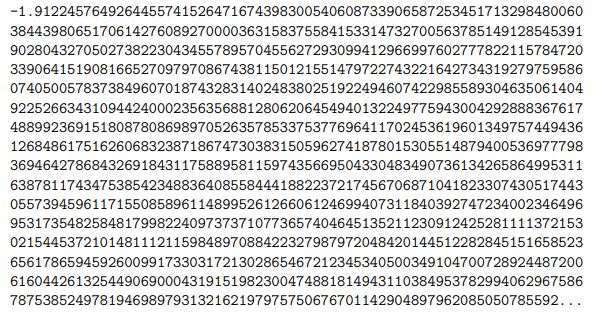}
\vspace{-10pt}
\caption{First 1100 digits of A8}
 \vspace{-10pt}
\end{figure}

In 2018, Kinoshita's team detected an error in his previous calculation of the tenth-order: \textbf{\textit{“we found that one of the integrals, called X024, was given a wrong value in the previous calculation due to an incorrect assignment of integration variables.”}} \cite{Kinoshita2018} The new value was +6.675. 

\begin{equation} 
	\begin{cases}
	& g =  1 + \dfrac {\alpha}{2 \pi} - 0.328 \left(\dfrac {\alpha}{\pi}\right)^2 + 1.181 \left(\dfrac {\alpha}{\pi}\right)^3 \\
	& - 1.912 \left(\dfrac {\alpha}{\pi}\right)^4  + 6.675 \left(\dfrac {\alpha}{\pi}\right)^5
	+  g \ (noQED)  \\
& g \ (theor) = 1.001,159,652,182,032(720) \\
& g \ (exp) = 1.001,159,652,180,73(28) \\
	\end{cases}
\end{equation}

For the twelfth time, the new theoretical value was in good agreement with the new experimental value.

\subsection{The muon anomaly}
Despite the enormous effort made in recent times, the discrepancy between the theoretical value and the experimental value of the muon g-factor could not be eliminated, maintaining an error greater than 3 sigmas. Theoretical physicists are concerned about this discrepancy, as it is perhaps the most palpable evidence that the Standard Model is incomplete.
\begin{equation} 
	\begin{cases}
& g \ (theor) =  1.001,165,918.04(51). \\
& g \ (exp) = 1.001,165,920.9(6) \\
	\end{cases}
\end{equation}

\begin{figure}[h]
\centering
\includegraphics[scale=0.45]{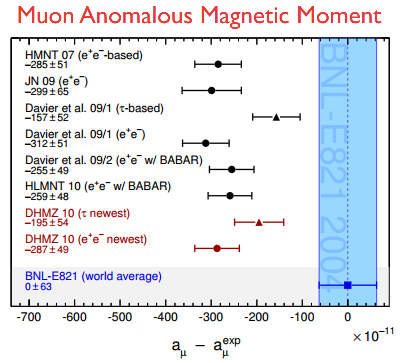}
\vspace{-10pt}
\caption{Muon anomaly}
\vspace{-5pt}
\end{figure}

In 2011, the E989 experiment was devised to improve the accuracy of the E821 experiment. This extremely complex experiment was performed at the Fermilab's Tevatron. Before the experiment could be conducted, a gigantic magnet (15 meters in diameter and 600 tons in weight) had to be moved 1300 km, from BNL to Fermilab. This delicate operation was successfully performed in June 2013. The magnet transfer lasted 35 days and cost 3 million dollars. 

\begin{figure}[h]
\centering
\includegraphics[scale=0.3]{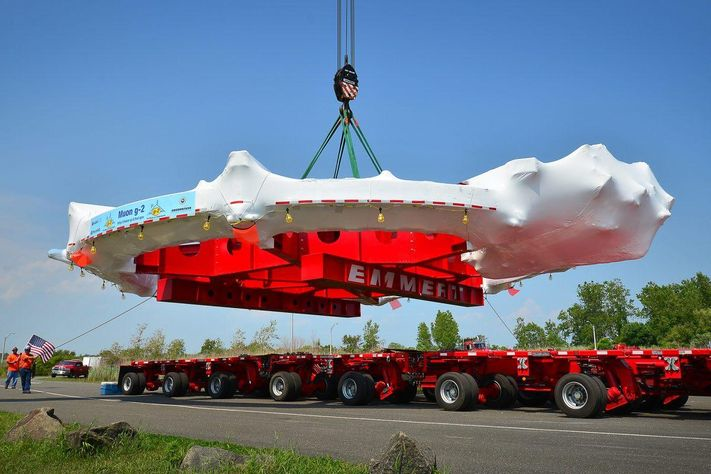}
\vspace{-10pt}
\caption{Transportation of the 600 ton magnet to Fermilab}
\vspace{-5pt}
\end{figure}

In addition, the Fermilab particle accelerator had to be enlarged. The related investment plan, the PIP-II Reference Design Report, had an estimated cost of 600 million dollars and was approved in July 2018. The results of E989 experiment were published in 2021, but they confirmed the anomaly once again.

\markright{\ptepauthor. \pteptitle}
\section{Summary}
\markright{\ptepauthor. \pteptitle}
According Feynman: \textbf{\textit{“We have found nothing wrong with the theory of quantum electrodynamics. It is, therefore, I would say, the jewel of physics; our proudest possession.”}} \cite{Feynmanbook}  But the reality of the QED is better reflected by Dyson’s \linebreak description in a letter to Gabrielse in 2006  \textbf{\textit{“As one of the inventors of QED, I remember that we thought of QED in 1949 as a temporary and jerry-built structure, with mathematical inconsistencies and renormalized infinities swept under the rug. We did not expect it to last more than 10 years before some more solidly built theory would replace it. Now, 57 years have gone by and that ramshackle structure still stands.”}}\cite{Dysonletter}

All calculations performed in QED always result in an infinite value. Renormalization techniques must be used to obtain finite results. These Renormalization techniques are not mathematically legitimate. Despite this fact, they have continued to be used because they provide results that fit perfectly with the experimental results. This has provided a kind of "empirical legitimacy" to renormalization.

But for this "empirical legitimacy" to be acceptable, there can be no doubt about the mathematical calculations used. Nevertheless, the lack of critical review of the theoretical results that have been published is evident. The theoretical results are only scrutinized when they do not match the experimental values. It is no longer a surprise that errors continually appear in theoretical calculations. Recall that the calculation of each Feynman diagram implies the resolution of multiple factors, and that each of these factors diverges to infinity. Therefore, renormalization techniques must be arbitrarily applied to eliminate these infinities and to obtain finite results. Moreover, these calculations are extremely complex and are not published in their entirety, so it is impossible to independently validate them.

\begin{flushright}\footnotesize
1 October 2021
\end{flushright}

\vspace*{-6pt}
\centerline{\rule{72pt}{0.4pt}}
}

\end{document}